\documentclass[11pt, draftclsnofoot, onecolumn]{IEEEtran}

\usepackage{cite,graphicx,amsmath,amssymb}
\usepackage{subfigure}
\usepackage{fancyhdr}
\usepackage{mdwmath}
\usepackage{mdwtab}
\usepackage{balance}
\usepackage{xcolor}
\usepackage{bm}
\usepackage{amsthm}
\usepackage{algorithm}
\usepackage{algorithmic}
\usepackage{multirow}
\usepackage{flafter}

\begin{document}
\title{UAV Communications Based on Non-Orthogonal Multiple Access}

\author{
\IEEEauthorblockN{Yuanwei Liu,
Zhijin Qin,
Yunlong Cai,
Yue Gao,
Geoffrey Ye Li,
and~Arumugam~Nallanathan
 }


\thanks{Y. Liu, Z. Qin, Y. Gao and A. Nallanathan are with Queen Mary University of London, London, UK, E1 4NS, email:\{yuanwei.liu, z.qin, yue.gao, a.nallanathan\}@qmul.ac.uk.}
\thanks{Y. Cai  is with Zhejiang University, Hangzhou, 310027, China, e-mail: ylcai@zju.edu.cn.}
\thanks{G. Y. Li is with Georgia Institute of Technology, Atlanta, GA, USA, 30332-0250, email: liye@ece.gatech.edu.}

}

\maketitle

\begin{abstract}

This article proposes a novel framework for unmaned aerial vehicle (UAV) networks with massive access capability supported by non-orthogonal multiple access (NOMA). In order to better understand NOMA enabled UAV networks, three case studies are carried out.  We first provide performance evaluation of NOMA enabled UAV networks by adopting stochastic geometry to model the positions of UAVs and ground users. Then we investigate the joint trajectory design and power allocation for static NOMA users based on a simplified two-dimensional (2D) model that UAV is flying around at fixed height. As a further advance, we demonstrate  the UAV placement issue with the aid of machine learning techniques when the ground users are roaming and the UAVs are capable of adjusting their positions in three-dimensions (3D) accordingly. With these case studies, we can comprehensively understand the UAV systems from fundamental theory to  practical implementation.

\end{abstract}

\section{Introduction}
With the rapid development of control technology and manufacture business, unmanned aerial vehicles (UAVs), which were originally sparked by the military use, have gradually demonstrated the civil potentials
to new applications and markets opportunities, such as advanced cargo distribution, aerial photography, and wildfire management, to name a few~\cite{UAV2008proceeding}.
Regarding communication areas, UAV-aided communication has been recognized as an emerging technique on both industry and academia for its superior on flexibility and autonomy. On the one hand, industry projects, such as Google Loon project,  Internet-delivery drone in Facebook \cite{Bennis2017TWC}, and airborne LTE services in AT$\&$T, have been deployed for providing airborne global massive connectivity. On the other hand, promising research scenarios for assisting 5G communications can be as follows: establishing temporal communication infrastructure during natural disasters, offloading traffic for dense networks, and data collection for supporting Internet of Things (IoT) networks, etc. \cite{Kobayashi:2014,Zeng2016WCM}.

On the road to integrate UAV to 5G networks and beyond, multiple access techniques are essential. Currently, non-orthogonal multiple access (NOMA) is  regarding as a promising candidate for 5G networks due to its superior on providing higher spectral efficiency and supporting massive connectivity \cite{Ding2015Application}. Therefore, NOMA has received significant attention on recent industry standardization, such as the 3rd Generation Partnership Project Long-Term Evolution Advanced (3GPP-LTE-A) standard, the 5G New Radio (NR) standard, and the next general digital TV standard (ATSC 3.0) \cite{Liu2017Proceeding}. The key concept of NOMA is to serve multiple users on the same resource block (time/frequency/code/space), but with different power levels. Particularly, as its name suggests, power-domain NOMA allocate the power resouces to multiplex users  with the aid of superposition coding (SC) at the transmitters. At the receivers, successive interference cancellation (SIC) is invoked for exploiting the channel difference of different users. As a consequence, NOMA is capable of enhancing connectivity and meeting diverse user requirements~\cite{Cai2017Survey}.

The goal of this article is to provide a potential solution to realize UAV networks with NOMA, spanning from the mathematical modeling, the joint resource and trajectory optimization  to aerial base station (BS) placement and movement design. Nevertheless, there are still several research challenges to be addressed, for example, the distinct characteristics of applying NOMA into UAV networks inevitably necessitate the re-design of user association algorithms. Moreover, considering the mobile features of UAV, NOMA is thereby capable of supporting dynamic user grouping and bringing more flexibility for the network design. As a consequence, more effective algorithms for dynamic three-dimension (3D) aerial BS placement and movement design are required to mitigate interference. All aforementioned challenges motivate us to contribute this article.

The rest of this article is organized as follows. First, we identify the key features of both NOMA and UAV networks in Section II. Then we discuss stochastic geometry based modelling for NOMA-aided UAV networks in Section III, followed by joint power allocation and trajectory design  in Section IV. Moreover, we investigate the use of machine learning to conduct dynamic aerial BS  placement and movement design within 3D space in Section V. Finally, conclusions and promising research directions are discussed in Section VI.

\section{Non-Orthogonal Multiple Access Aided UAV Networks}
Before introducing the UAV networks with NOMA, we characterize the unique features of UAV networks first. Generally, UAV networks have the following characteristics:
\begin{itemize}
  \item \textbf{Path loss}: Since there are usually not many obstacles in the air, we use a simplified model to assume that the line-of-sight (LOS) links between  the UAVs and the users are dominated, which are significantly less effected by shadowing and fading. In more complicated practical scenarios, such as urban areas where buildings and other obstacles on the ground may block UAV flight and signal transmission, both LOS and non-line-of-sight (NLOS) links require to be considered.

  \item \textbf{Mobility}: When a UAV flies around, the coverage areas becomes various. Therefore, the UAV can support different ground users. For example, UAVs are capable of roaming above a group of users to enhance the channel conditions so as to provide high throughput.

    \item \textbf{Agility}: Based on the real-time requirements from the users, UAVs can be deployed quickly and  their positions can be adjusted within a 3D space flexibly, which enables UAV networks to provide flexible and on-demand service to the ground users with lower costs compared to the terrestrial BS.
\end{itemize}

\begin{figure}[t!]
    \begin{center}
        \includegraphics[width=6in]{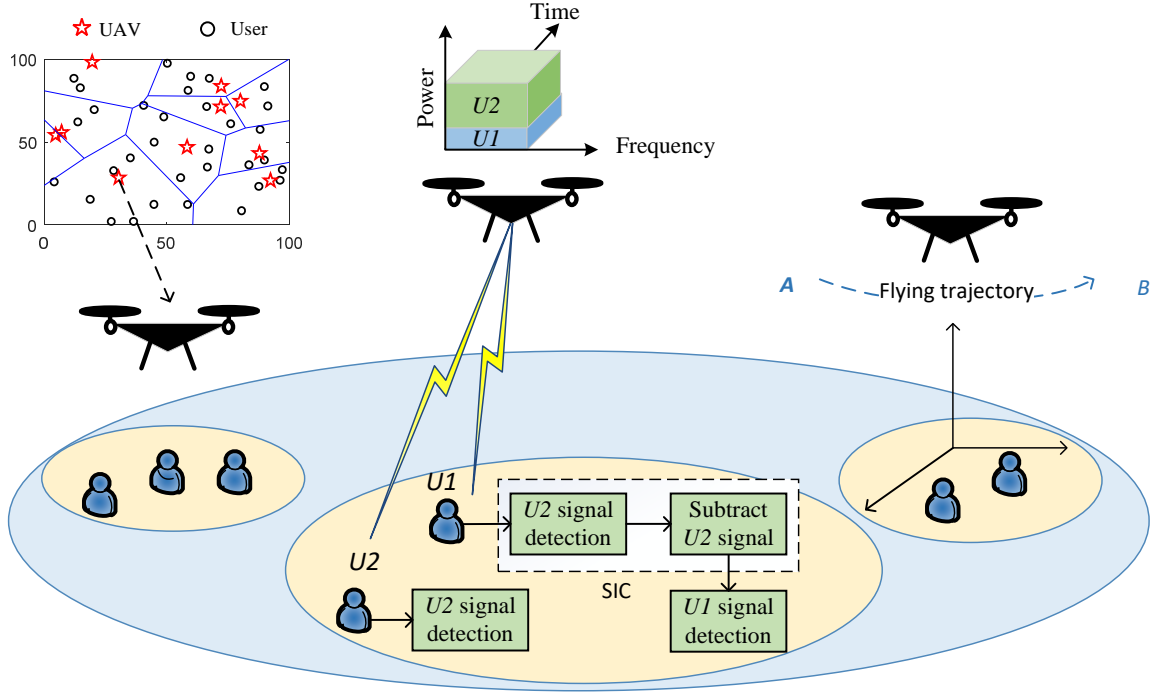}
        \caption{Illustration for NOMA-aided UAV networks.}
        \label{framework}
    \end{center}
\end{figure}

To support massive connectivity in UAV networks, power-domain NOMA can be adopted to support different users over the same time/frequency slot based on the aforementioned three features. Note that UAVs communicate with ground users mainly relying on  LoS links. Such characteristics make that there are no distinct channel gain differences between UAVs to multiple NOMA ground users compared to conventional terrestrial NOMA communications. Therefore, the NOMA-aided UAV networks should be carefully re-designed according to large-scale fading differences of NOMA users by invoking user pairing/grouping techniques, which will be detailed in Section III. Fig.~\ref{framework} shows some scenarios in the downlink of NOMA-aided UAV networks. By adopting the NOMA technique in the downlink transmission, different users can share the same resource block by using different transmit power levels with a total power constraint. Taking the case shown in the middle of Fig.~\ref{framework} as an example, where there \textcolor[rgb]{0.00,0.00,1.00}{is} a UAV with two covered users, i.e., $U1$ and $U2$,  the UAV  sends a superimposed signal containing the signals for the two users. In order to guarantee the fairness, less power is allocated to  $U1$ that is with better downlink channel state information (CSI). SIC is used for signal detection at the receiver. $U2$, with higher transmit power and poorer channel gain, is decoded first by treating $U1$ as  interference. Once $U2$ is detected and decoded, its signal component will be subtracted from the received signal to facilitate the detection of $U1$. By doing so, $U2$ suffers from the higher inter-user interference and the detection error of $U2$ will pass to $U1$. Therefore, we have to allocate sufficient power to the user that is detected first, i.e., $U2$.

The logic behind Fig.~\ref{framework} is to provide a comprehensive understanding of the design of practical NOMA-aided UAV networks. On the left side of Fig.~\ref{framework}, the Voronoi figure shows the spatial relationship of UAVs and NOMA users, where the five-pointed stars representing the UAVs are the vertical mapping from the air to the ground.  The right side of Fig.~\ref{framework} shows the trajectory of a UAV when it serves the ground users. Fig.~\ref{framework} as a whole presents a complicated system with multiple UAVs and dynamic  users. Based on the three characteristics and new requirements of various scenarios in NOMA-aided UAV networks, the following issues should be addressed accordingly:
\begin{itemize}
\item \textbf{Random spatial modeling:} To quantify NOMA-aided UAV networks with LOS and Nakagami-m fading channels, performance analysis of NOMA-aided networks is desired before designing and implementing such  networks. Scholastic geometry is an effective tool to analyze the average performance of the networks.  Using a stochastic geometry based model, the ground users are grouped together to share the same resource with NOMA and associated to the most suitable UAV. Therefore,  we will investigate users grouping and association of NOMA-aided UAV networks in Section III.

  \item \textbf{Resource allocation and trajectory design}: By considering the mobility of UAVs, trajectory design is necessary to provide better coverage for various ground users. As shown on the right side of Fig.~\ref{framework}, from the starting point, $A$, to the destination, $B$, the power allocation of the NOMA users should be addressed properly to further improve the system performance. We present a joint optimization of resource allocation and trajectory design with LOS links in Section IV.
  \item \textbf{Dynamic movement and deployment design}: When there are multiple UAVs in the networks and the NOMA users are roaming dynamically, how to adjust the positions of UAVs dynamically to optimize the system performance with considering dynamic user grouping of NOMA users becomes more challenging, especially when there are both the LOS and NLOS links between transmitters and receivers. The  movement and deployment of multiple UAVs should be carefully designed in the dynamic environment. To address these complicated issues, we present a novel framework based on machine learning in Section V.
\end{itemize}

\section{NOMA-aided UAV Networks}

It is worth analyzing the average performance of NOMA-aided UAV networks in order to obtain important insights for the network design and implementation.  Stochastic geometry is a powerful tool to capture the spatial randomness of wireless networks. In this section, we propose a new UAV framework for randomly roaming NOMA users. Particularly, we consider a two-tier spatial model that  NOMA users are served by UAVs at a fixed altitude. Such spatial model can be extended to the case that UAVs fly with varied altitudes. As mentioned in Section II, the LOS links are dominant in UAV nertworks, conventional Rayleigh fading becomes improper for representing the channel characteristics. We apply Nakagami-m fading  to model the small-scale fading in NOMA-aided UAV networks. 

\subsection{Single-UAV case}

\begin{figure}[t!]
    \begin{center}
        \includegraphics[width=4.0in]{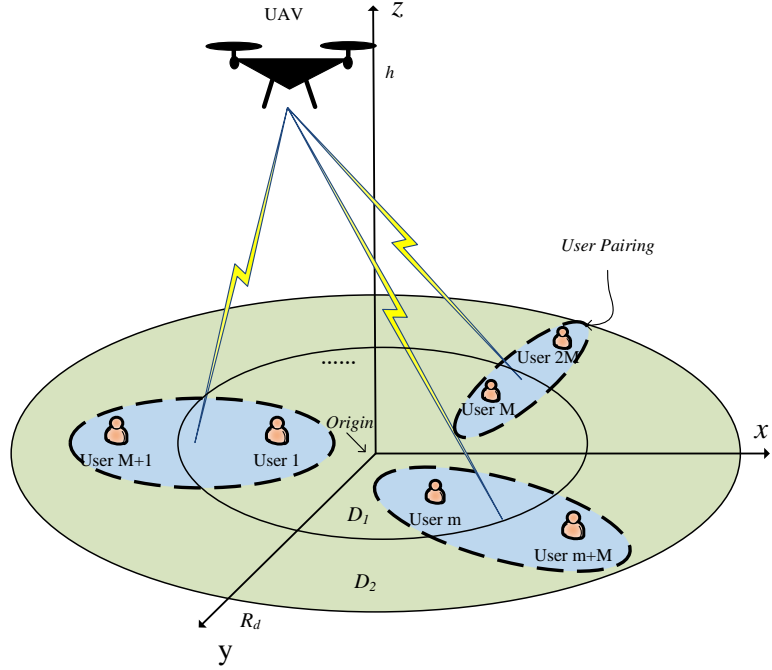}
        \caption{Illustration of user pairing in NOMA-aided UAV networks.}
        \label{user_pair}
    \end{center}
\end{figure}

For the single-UAV case, as shown in Fig.~\ref{user_pair}, the cell coverage of UAV is a disc area $D$, with a radius of $R_d$. The UAV that communicates with $2M$ users is located at the center of the disc, with a vertical height $h$. As NOMA requires user pairing, the $2M$ users are divided into $M$ orthogonal pairs first. Each pair is then randomly allocated to an orthogonal resource block.  To generate more distinct channel quality differences between the paired user, we divide the disc $D$ into two regions, $D_1$ and $D_2$, for the group of cell-center users and the group of cell-edge users\footnote{Cell-center users refer to users that are close to UAVs while cell-edge users refer to users that are far from UAVs.}, respectively. For each paired user, the cell-center user is assumed to be capable of cancelling the interference of the cell-edge user by the SIC technique. In \cite{Hou2018TCOM}, a multiple antenna aided NOMA has been considered in UAV networks by utilizing such a spatial system with applying outage probability and ergodic rate as performance metrics. Numerical results in \cite{Hou2018TCOM} reveal the performance gain of the proposed NOMA-aided UAV framework over the conventional orthogonal multiple access (OMA) framework.

As we distinguish the two groups of users according to their distances to the connected UAVs, a critical question arises here: which cell-center NOMA user should be paired with which cell-edge user? There are three possible user pairing strategies based on the locations of users \cite{Yuanwei2016JSAC}:
\begin{enumerate}
  \item  A cell-center user is randomly paired with a cell-edge user, which is the benchmark strategy.  This strategy does not require full CSI of all users.
  \item  The nearest cell-center user, which has the closest distance to the connected UAV among the group of cell-center users is paired with the nearest cell-edge user, which has the closest distance to the the connected UAV among the group of cell-edge users. Such strategy has the best system performance.
  \item  The nearest cell-center user among the group of cell-center users is paired with the farthest cell-edge user  among the group of cell-edge users. Such strategy  provides the largest NOMA gain over the conventional OMA scheme.
\end{enumerate}

Nevertheless, all of the aforementioned user pairing strategies are for the sake of enhancing the performance of single pair, which does not necessarily improve the performance of the NOMA-aided UAV networks with multiple users. The multiple-pair network performance with considering resource management among different resource blocks requires further investigation.

\subsection{Multiple-UAV case}

Compared to the single-UAV case, the NOMA-aided multiple-UAV case is more challenging to analyze.  We consider a large-scale NOMA-aided UAV networks where the UAVs and NOMA users are modeled by two independent homogeneous poisson point processes (HPPPs). Specifically, the NOMA users are deployed in an infinite  horizontal plane with density $\lambda_u$ while the UAVs are deployed in another infinite plane with a fixed altitude $h$ and density $\lambda_v$. Each UAV is supposed to communicate with $K$ NOMA users. As we have multiple UAVs and a number of users and the distributions of the UAVs and the users are independent from each other, a flexible yet simple user association policy is required. Three possible user association policies are detailed in the following.

The first possible user association policy is based on the distances, i.e., one UAV is associated with the $K$ nearest NOMA users. This is the simplest user association policy. However, such user association policy is not flexible as different UAVs may have different features, such as transmit power, multiple-antenna aided beamforming gain, etc. To solve this problem, a more flexible user association scheme based on the average received power can be invoked. However, the average received power of each NOMA user is not predetermined and dependents on the power allocation scheme. As a result, the power allocation among NOMA users is very important and can further enhance the network performance. Taking into considerations of mobility and agility features of UAVs mentioned in Section II, the third promising user association policy is based on the maximum instantaneous signal-to-interference-plus-noise (SINR) of each NOMA user. Nonetheless, such user association policy requires instantaneous CSI at both UAVs and NOMA users, and therefore incurs  heavy overhead.

\section{Joint Power Allocation and Trajectory Design}
After analyzing the average performance of UAV networks with NOMA, we focus our attentions on the design of UAV's trajectory. In conventional cellular networks, in order to support more users or achieve higher data rates, more small cell BSs are installed. While in UAV networks, with the high mobility of UAVs, the communication distance between the UAV and ground users can be adjusted based on the real-time requirements. By carefully designing the trajectories of UAVs,  better coverage can be achieved. The SIC order, determined by the received signal strengthes of difference users, varies with the locations of UAVs. Therefore, we need to jointly consider the transmit power of different NOMA users and   trajectory of the UAV to ensure that the transmit signals can be decoded accurately. In this section, we provide some initial trials on the joint trajectory design and power allocation in a single UAV network with NOMA.

As shown in the right side of Fig.~\ref{framework}, a UAV is flying from the initial point, $A$, to the destination, $B$, with the same height during the flying time $T$. It can provide service to  several ground users in  the downlink transmission. There have been some initial research contributions in the context of the joint optimization of trajectory design and resource allocation for UAV networks  with OMA~\cite{Zeng:2017:TWC,Lyu:2018:TWC,Xu:2018:TWC}. In order to provide more access opportunities, the UAV adopts power-domain NOMA to serve those ground users within the same channel simultaneously~\cite{Cui:TWC:2018}. In the considered network, the UAV acts as a transmitter that sends a superimposed signal to all the ground users. The channels between UAV and  ground users are dominated by LOS links. At the receivers, the signals for different users are detected via SIC in the descending order of channel gains. It is noted that the trajectory of the UAV  influences the achievable network performance. Moreover, as the SIC decoding order of the ground users keeps changing during the flying, the power allocated to different users should be adjusted accordingly. Therefore, a joint power and trajectory design is desired in UAV networks with NOMA to maximize the minimum average rate of the NOMA users under the constraints of the UAV's speed, transmit power, initial and final locations, and the SIC order.

\begin{figure*}[t]
    \centering
    \subfigure[Max-min rate comparison of OMA and NOMA in UAV networks with optimized trajectory.]{\label{max_min_rate}
\includegraphics[width= 3.0in,height=2.3in]{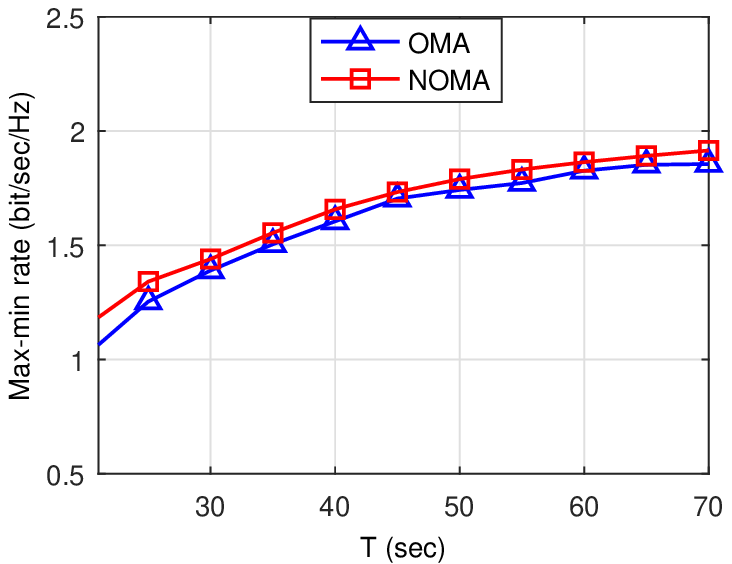}}
    \subfigure[Optimized UAV trajectories for  T = 25 seconds.]{\label{T25}
\includegraphics[width= 3.2in,height=2.3in]{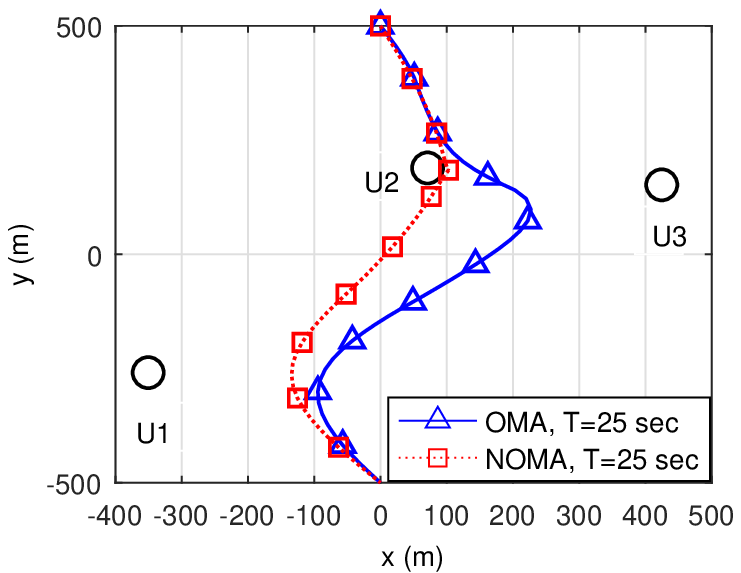}}
    \caption{Illustrations of UAV networks where three users that are randomly and uniformly distributed within an area of $1 \times 1 km^2$. Flight route is from initial location ${\left[ {0,500} \right]^T}$ to the destination ${\left[ {0, - 500} \right]^T}$ with the altitude 100 m.~\cite{Cui:TWC:2018}}
    \label{approx_simulation}
\end{figure*}

Fig.~\ref{max_min_rate} compares the performance  of UAV networks with OMA and NOMA, respectively, where a UAV is adopted to serve the ground users. In the OMA-aided UAV networks, the ground users are served sequentially at different time slots. With different flight durations, $T$,  the performance of the NOMA-aided UAV networks always outperforms that of the OMA-aided UAV networks in terms of max-min rate of all served users, which indicates that NOMA should be used in UAV networks. Moreover,  performance of the NOMA-aided UAV networks can be further improved when the trajectory and user scheduling are carefully designed. Fig.~\ref{T25} shows the corresponding trajectory of the UAV during the flying period of $T=25$ seconds. From the figure, the trajectories for the NOMA mode and the OMA mode are quite different. In the OMA mode, the UAV intends to get close to each user successively for fairness. Since the UAV is close to $U2$, more power is allocated to the remote users, e.g., $U1$ and $U3$, for fairness. From the density of the samples over the trajectories, we can also see that the UAV flies at a lower speed when it gets close to $U2$ in the NOMA mode.

\section{Dynamic UAV Placement and Movement Design: A Machine Learning Approach}

In previous two sections, we assume that the UAV is floating on a plane with a fixed altitude. Actually, if we consider the simplified model with only LOS links between UAVs and NOMA users, UAVs should fly as low as possible to minimize the serving distances. However, some scenarios that are topographically sophisticated, such as  in urban or mountain areas, there exist both  LOS and NLOS links \cite{Al-Hourani:2014}. Moreover, decreasing the altitudes of UAVs results in lower path loss but lower LOS probability as well. Therefore, the altitudes of UAVs also require to be dynamically adjusted, which makes trajectory design more complicated. In this section, we will introduce a novel framework for dynamically designing placement and movement for the NOMA enhanced multiple UAVs flying in the 3D space. 

In contract to the conventional trajectory design where there is a fixed starting point and destination for a dedicated UAV,  the movement design is based on the real-time user requirements. For example, for the area that users are densely deployed and require massive connectivity, we can divide the area into different clusters. We apply a novel hybrid multiple access technique where NOMA is applied in the same cluster and time division multiple access (TDMA) is applied among different clusters. UAVs are moving over those clusters in different time slots accordingly.  When the group of users require high data rates, i.e., uploading/downloading high dimension  videos, we can adjust the altitudes of UAVs for enhancing the channel qualities. A more practical example is that ground users are randomly roaming in reality, the number of NOMA users in each group varies. Therefore, the positions of UAVs should also be adjusted accordingly in real time.
\begin{figure}[t!]
    \begin{center}
        \includegraphics[width=4.5in]{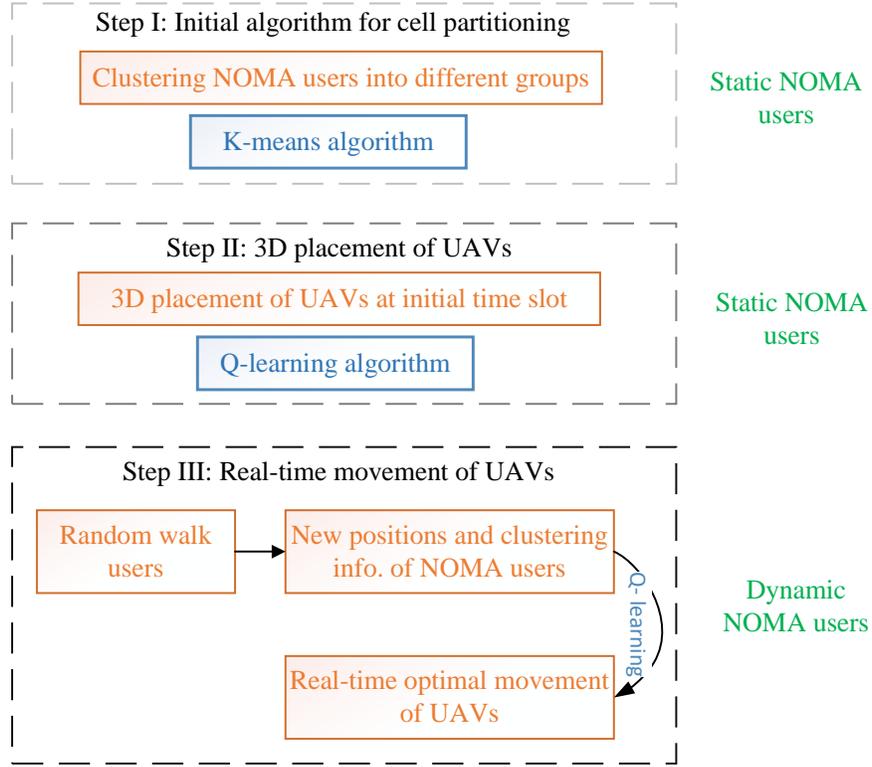}
        \caption{Three-step machine learning based UAV placement and movement design in NOMA-aided networks.}
        \label{QL framework}
    \end{center}
\end{figure}

As shown in Fig.~\ref{QL framework}, we propose a three-step machine learning based framework to solve the dynamic 3D UAV placement and movement design problem, the first step is to invoke the K-means algorithm to obtain the initial cell positioning of the ground users and the number of NOMA users in each cell. The second step is using the Q-learning-based \emph{placement} algorithm for 3D placement of UAVs by considering the users as static. The third step is using a Q-learning based \emph{movement} algorithm to obtain 3D real-time movement of UAVs where the users are following random walk model. It is worth noting that the proposed framework contains the training and testing process, which is capable of training the policy for making decision offline. By doing so, the complexity can be reduced significantly.

\begin{figure}[t!]
    \begin{center}
        \includegraphics[width=3.8in]{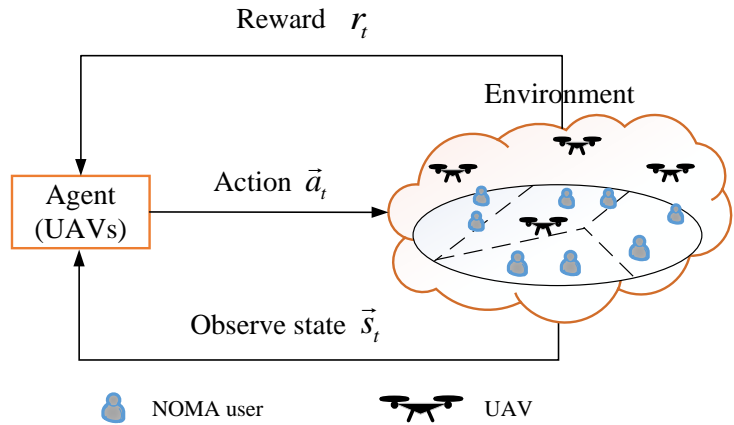}
        \caption{The procedures of Q-learning in the UAV movement design.}
        \label{reward}
    \end{center}
\end{figure}

Note that the core part of the proposed framework is Q-learning as shown in Fig. \ref{reward}, the basic operating principle of Q-learning is as follows. A basic Q-learning model consists of four parts: states of agents, actions, rewards, and Q-value (on policy). The aim of Q-learning is to obtain a policy (a set of actions) that maximizes the observed rewards over the interaction time~\cite{DeepLearning:SPM}. Our proposed framework, mainly contains three steps as follows:

\begin{enumerate}
  \item At each time slot $t$, each UAV observes one state  from the state space vector ${\vec {\mathbf{s}}_t} = \left[ {{s_{1,t}},{s_{2,t}}...,{s_{N,t}}} \right]$, which records the coordinate information of all UAVs, where ${s_{n,t}}$ is the 3D coordinate information for the $n$-th UAV, at time $t$.
  \item The $n$-th UAV takes an action ${a_{n,t}}$ accordingly from the action space ${\vec {\mathbf{a}}_t} = \left[ {{a_{1,t}},{a_{2,t}}...,{a_{N,t}}} \right]$, based on the current state space vector, the directions of UAVs based on the decision policy. For example, ${a_{n,t}}= (0,0,1)$ represents that the $n$-th UAV moves one step to the up direction. Such decision policy is determined by a Q-table, Q (${\vec {\mathbf{s}}_t}$, ${\vec {\mathbf{a}}_t}$). The decision policy is to select actions to obtain maximum Q-value during each time slot.
  \item Following the action, the state of the $n$-th UAV switch to a new state $S_{n,t+1}$. An instantaneous reward, $r_t$, is obtained as a results of the action.
\end{enumerate}

The three steps are repeated  to adjust the placement and movement of UAVs.

\section{Future Challenges and Conclusion Remarks}
In this article, the design challenges of integrating NOMA techniques into UAV networks have been investigated. The key features of UAVs have been characterized first, which enable new opportunities for NOMA to provide massive connectivity. Then NOMA-aided UAV networks have been modelled and analyzed with the aid of stochastic geometry, for both single-UAV and multiple-UAV cases. This is followed by introducing the joint power allocation and trajectory design for NOMA-aided single-UAV case. Finally, a machine learning framework has been proposed to solve the dynamic placement and movement of UAVs in the 3D space. However, there still exist some open research issues in context of the implementation of NOMA enabled UAV networks,  which  are highlighted as follows:
\begin{itemize}
\item \textbf{A unified spatial model for NOMA-aided UAV networks}: there are various communication scenarios for NOMA-aided UAV networks, e.g., single-UAV case, multiple-UAV case, uplink, downlink, cooperative communications scenarios, etc. A unified spatial analytical framework for NOMA-aided UAV networks is desired, which can be effortless switched to fit different practical application scenarios.

  \item \textbf{Data driven NOMA-aided UAV networks design}: Big data has been recognized as a powerful tool to provide insightful guidelines for real systems. Most of current research contributions in context of NOMA-aided UAV networks are based on data generated randomly, which may be different from practical situations. Data from social networks, such as Twitter and Facebook, can be used for collecting position information of mobile users. As a further advance, more sophisticated big data analytical approaches, such as data mining and stochastic modeling, can be invoked for analyzing the historical data and providing more accurate prediction in terms of NOMA users' mobilities. By doing so, the UAVs are able to adjust their positions more accurately to further improve the system performance.

  \item \textbf{MIMO-NOMA design in UAV networks}: NOMA is expected to co-exist with MIMO techniques for further improving the spectral efficiency and supporting massive connectivity of UAV networks. Nevertheless, applying multiple antenna techniques in NOMA requires carefully designing the channel ordering. In contract to the single antenna NOMA case in which the channels of users are scalers; those in multiple antenna aided NOMA are vectors or matrix. Additionally, due to the 3D characteristics of UAV networks, beamforming based or cluster based MIMO-NOMA design becomes more challenging. As a result, how to order channels in MIMO-NOMA systems with considering the characteristics of UAV networks requires more research contributions.
  \item \textbf{Low latency design for NOMA-aided UAV networks}: It is worth pointing out that one remarkable feature of UAVs is agility. The SIC decoding characteristics of NOMA inevitably brings considerable delays at receivers if the number of NOMA users is large. A possible solution is to adopt hybrid multiple access by dividing a large number of NOMA users into different orthogonal groups. In each group, a small number of users invoke NOMA for decreasing the delay caused by SIC.
\end{itemize}

\bibliographystyle{IEEEtran}

 \end{document}